# HUMAN FACTORS OF FORMAL METHODS

Maria Spichkova

*Institut für Informatik, Technische Universität München*
*Boltzmannstr. 3, 85748 Garching, Germany*

**ABSTRACT**

This paper provides a brief introduction to the work that aims to apply the achievements within the area of engineering psychology to the area of formal methods, focusing on the specification phase of a system development process.

## 1. INTRODUCTION

There are many definitions of human factors, however most of them (applying to the field of software and system engineering) are solely oriented on human-machine system operation in terms of system and program usability, i.e. on those parts that are seen by the (end-)user, but not by the requirements, specification and verification engineers. The fundamental goal of human factors engineering (see also [Wickens, Hollands 2000]) is to reduce errors, increase productivity and safety when the human interacts with a system. Engineering psychology applies psychological perspective to the problems of system design and focuses on the information-processing capacities of the human brain too.

There are many applications of formal methods to analyze human-machine interaction and to construct user interfaces (see, e.g., [Shackel, Richardson 1991]), but the field of application of human factors to the analysis and to the optimization of formal methods area is almost unexplored. To our best knowledge there are no other works on this field, the nearest area is only the application of human factors to the development of engineering tools. Dealing with formal methods it is often assumed that only two factors must be satisfied: the method must be sound and give such a representation, which is short and beautiful from the mathematical point of view. This leads to the fact that formal methods are treated by most engineers as "something that is theoretically important but practically too hard to understand and to use", where even small changes of a formal method can make it more understandable and usable for an average engineer.

Formal methods are especially important for the development of safety-critical systems, and speaking about human factors according to this kind of systems we focus mostly on technical aspects (so called *Engineering Error Paradigm*, see also [Redmill, Rajan 1997]) in application to the formal methods. Human factors that are targeted by the Engineering Error Paradigm typically include the design of man-machine interfaces as well as the corresponding automatization: by this paradigm humans are seen as they are almost equivalent to software and hardware components in the sense of operation with data and other components, but at the same time humans are seen as "the most unreliable component" of the total system. This implies also that designing humans out of the main system actions through automatization of some system steps (in our case, automatic translation from one representation kind to another one, translation between formal languages etc.) is considered as a proposal for reducing risk. Another important view of this paradigm is that human errors often occur as a result of mismatch in human-machine interface, overestimation of physical capabilities of a person – human performance and reliability (in our case, e.g., clearness – up to obviousness – and readability of formal specification) need to be considered in the design process.

In our approach *Human Factors of Formal Methods*, HF$^2$M, we focus on human factors in formal methods used within formal specification phase of a system development process: on (formal) requirements specification and on the developing of a system architecture that builds a bridge between requirements and the corresponding system.

## 2. HUMAN FACTORS + FORMAL METHODS = HF$^2$M

As mentioned in our previous work [Spichkova 2007], during requirements specification phase and the phase of a system architecture development phases we need to care about later phases (modelling, simulation, testing, formal verification, implementation) already doing the formal or, even, semiformal specification of a system – that is, choosing an appropriate abstraction and modelling technique. A crucial question is here how we can optimize the formal representation and formal methods with respect to human factors.

The main ideas of the approach are language/framework independent, but for a better readability and for better understanding of these ideas we show them base on formal specifications presented in the Focus (see also [Broy, Stølen 2001]), a framework for formal specifications and development of interactive systems.[1] We can also see this methodology as an extension of the approach "Focus on Isabelle" [Spichkova 2007] – it is integrated into a seamless development process, which covers both specification and verification, starts from informal specification and finishes by the corresponding verified C code.

### 2.1 Specification

The main aspect of HF$^2$M is the representation (layout and visualization including graphical representation) of formal specification. This work was started within the Verisoft-XT project (see [Hölzl et al. 2010], [Spichkova et al. 2012]), however, to get solid results, a well-founded study of current research within engineering psychology area as well as its application to formal methods (in general and based on the example of the Focus specification language) is needed.

The first results of visual optimization of Focus specifications are presented in [Spichkova 2011b]. This covers all specification styles – from textual representation to state transition diagrams (Figure 1 shows an example of such an optimisation where the difference in readability is seen immediately, also without understanding the meaning of concrete transition and states) and timed tables. Inter alia, we suggest to simplify the timed specification in the following way to get shorter specifications that are more readable and clear: specifying a component we have often such a case where for some time intervals both conditions hold: local variables are still unchanged and there is no output. This can occur, e.g., if at this time interval the component gets no input or if some preconditions don't hold. In classical Focus, as well as in Isabelle, we need to specify such cases explicitly otherwise we get an underspecified component that has no information how to act in these cases.

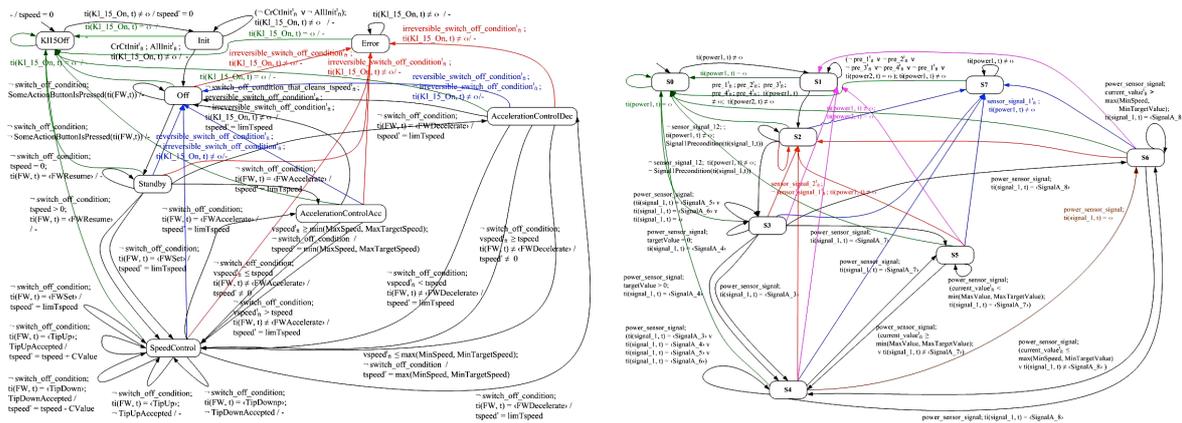

Figure 1. State transition diagrams: before and after optimization

---

[1] See http://focus.in.tum.de.

In many cases even not very complicated optimization changes of a specification method can make it more understandable and usable. Such a simple kind of optimization is often overlooked just because of its obviousness, and it would be wrong to ignore the possibility to optimize the language without much effort. For example, simply adding an enumeration to the formulas in a large formal specification makes its validation on the level of specification and discussion with co-operating experts much easier.

## 2.2 Decomposition

Another important aspect is a (formal) decomposition of formal specification: to cover it we introduced a number of decomposition methods, which are applied not only to keep the specification readable and manageable but also find inconsistencies and underspecification already during specification phase, without starting a formal verification process (see [Spichkova 2010], [Spichkova 2011a]).

The main difference and the main contribution of our methodology is that it was developed for such a system architecture where we have already specified systems or components properties in a formal way and need to decompose this whole properties collection to a number of subcomponents to get readable and manageable specifications. Thus, the presented methodology allows us to decompose system or component architecture exactly on the point where we see that the component specification becomes too large and too complex.

## 2.3 Add-ons

Covering different application areas by a single language helps to simplify representation of different views on a system as well as to switch between them. For this purpose we made the following extension/optimization of formal language for the Focus language:
- Specification of processes and matching to the representation of components,
- Specification of security-critical systems with respect to secrecy properties.

Specifying components and system in a formal language is helpful to have a possibility to change the view on the system or the kind of its description to cover some problem areas. For these reasons we extend the formal language Focus by the theory of processes described in [Leuxner 2010]. A process is understood there as "*an observable activity executed by one or several actors, which might be persons, components, technical systems, or combinations thereof*". Each process has one *entry (activation, start) point* and one *exit (end) point*. An entry point is a special kind of input signal/channel that activates the process, while an exit point is a special kind of output signal/channel that is used to indicate that the process is finished. We treat a process as a special kind of a Focus component having additionally two channels (one input and one output channel) of a special kind. These channels represent the entry and exit points of the process. All the technical details are presented in [Spichkova 2011b].

Dealing with *security-critical systems* we have an other question in the foreground: how we can combine system components that each enforce a particular security requirement in a way that allows us to predict which properties the combined system will have. For this purpose we introduced in [Spichkova 2012] a representation methodology for crypto-based software, such as cryptographic protocols, and their composition properties. Having such a formal representation, one can argue about the protocol properties as well as the composition properties of different cryptographic protocols in a methodological way and make a formal proof of them using a theorem prover.

Last but not least point is an appropriate automatization of some steps of formal specification and verification, because the automatization not only saves time but also excludes human element (as the most "unreliable") in failure. For this purpose we plan to extend the AutoFocus Case Tool (a scientific prototype[2] implementing a modelling language based on a graphical notation and the formal Focus semantics, see also [Huber et al. 1996]) by a number of add-ons, e.g., by add-on than allows tool-support of the methodology "*Focus on Isabelle*" (cf. [Spichkova 2007]) – a translator of formal specifications in Focus to Isabelle/HOL.

---

[2] http://af3.in.tum.de

## 3. CONCLUSION

In our work "*Human Factors of Formal Methods*" we aim to apply the achievements with the area of engineering psychology to the area of formal methods, focusing on the specification phase of a system development process. According to the Engineering Error Paradigm we optimize representation of formal specification (corresponds to the classical "design of man-machine interfaces") as well as add a corresponding automatization of some steps of formal specification and verification. We can apply our approach for making abstract views, decomposition, logical architecture development, corresponding refinement, etc. The main ideas of the approach are language/framework independent, but for better readability and for better understanding of these ideas we show them on the base of formal specifications presented in the Focus specification framework.